\documentclass[preprint,nofootinbib,showpacs,aps,prc]{revtex4-1}
\usepackage{graphicx}
\usepackage{amstext}
\usepackage{amssymb}

\usepackage[usenames]{color}
\begin{document}
\title{Femtoscopic scales in $p + p$ and $p+$Pb collisions in view of the uncertainty principle}

\author{V.M. Shapoval$^a$, P. Braun-Munzinger$^{b,c}$, Iu.A. Karpenko$^{a,c}$  }
\author{Yu.M. Sinyukov$^{a}$}
\affiliation{$(a)$ Bogolyubov Institute for Theoretical Physics,
  Metrolohichna str. 14b, 03680 Kiev, Ukraine \\ $(b)$ ExtreMe Matter
  Institute EMMI, GSI~Helmholtz~Zentrum f\"ur~Schwerionenforschung, 
D-64291 Darmstadt, Germany\\ $(c)$ Frankfurt~Institute~for~Advanced~Studies, Ruth-Moufang-Str.~1, 60438 Frankfurt am Main, Germany}

\begin{abstract}
  A method for quantum corrections of Hanbury-Brown/Twiss (HBT)
  interferometric radii produced by semi-classical event generators is
  proposed. These corrections account for the basic
  indistinguishability and mutual coherence of closely located
  emitters caused by the uncertainty principle. A detailed analysis is
  presented for pion interferometry in $p+p$ collisions at LHC energy
  ($\sqrt{s}=7$ TeV). A prediction is also presented of pion
  interferometric radii for $p+$Pb collisions at $\sqrt{s}=5.02$
  TeV. The hydrodynamic/hydrokinetic model with UrQMD cascade as
  'afterburner' is utilized for this aim.  It is found that quantum
  corrections to the interferometry radii improve significantly the
  event generator results which typically overestimate the
  experimental radii of small systems. A successful
  description of the interferometry structure of $p+p$ collisions
  within the corrected hydrodynamic model requires the study of the
  problem of thermalization mechanism, still a  fundamental issue
  for ultrarelativistic $A+A$ collisions, also for high multiplicity
  $p+p$ and $p+$Pb events.
\end{abstract}

 \maketitle


%Corresponding author: {\small \textit{Yu.M. Sinyukov, Bogolyubov Institute
%for Theoretical Physics, Kiev 03680, Metrolohichna 14b, Ukraine. E-mail:
%sinyukov@bitp.kiev.ua}}

\section{Introduction}
The quantum-statistical enhancement of the pairs of identical pions
produced with close momenta  was observed first in $\bar{p}+p$
collisions in 1959 \cite{Gold}. It took more than a decade to develop
the method of pion interferometry  based on the
discovered phenomenon. This was done at the beginning of the 1970s by
Kopylov and Podgoretsky \cite{Kopylov}. Their theoretical analysis
assumed the radiating source as consisting of independent incoherent
emitters. In fact, such a representation is used for a long time
for the analysis of the space-time structure of particle sources
created in $\bar{p}+p$, $p+p$, $e^+ + e^-$ and $A+A$ collisions.
The concept of independent emitters was applied to a further
development of the interferometric method, in particular, to account for
momentum-position correlations of the emitted particles \
\cite{Pratt, MakSin, Sinyuk, Hama} that, in turn, has resulted in
a general interpretation of the measured radii as the
homogeneity lengths in the Wigner functions \cite{Sin, AkkSin,
SinAkkKarp}. This concept is important for a study of
$A+A$ collision processes  within the hydrodynamic approach. Also a
detailed analysis of the particle final state (Coulomb) interactions
brings the significant contribution to the traditional method of
correlation femtoscopy \cite{FSI,baym}.

In a recent paper \cite{SinShap} the correlation 
analysis is taken  beyond the model of independent particle emitters.
It is found that the uncertainty principle leads to (partial)
indistinguishability of closely located emitters that fundamentally
impedes their full independence and incoherence. The partial
coherence of emitted particles is because of the quantum nature of particle
emission and happens even if there is no specific mechanism
to produce a coherent component of the source radiation. This
effect leads to a reduction of the interferometry radii and
suppression of the Bose-Einstein correlation functions. The effect
is significant only for small sources with typical sizes less than 2
fm. We shall apply this approach \cite{SinShap} to the
analysis of data in $p+p$ collisions at the LHC
energy of  $\sqrt{s}=7$ TeV, where the measured interferometry radii are
just within the above scale. A simple estimate will be done also for
$p+$Pb,  where the radii are larger and such corrections are less important.

A first attempt of the systematic theoretical analysis of the pion
interferometry of $p+p$ collisions at the top RHIC and $\sqrt{s}=0.9$
TeV LHC energies was made in Ref.\cite{QGSM} within the quark-gluon
string model (QGSM). It was found that, for a satisfactory
description of the interferometry radii, one needs to reduce
significantly the formation time by increasing the string tension
value relative to the one fixed by the QGSM description of
the spectra and multiplicity. Otherwise, the radii obtained within QGSM
are too large compared to the measured ones. The similar result is obtained within UrQMD \cite{Marcus}.
Hypothetically one can hope to reduce the predicted radii suggesting
the other approach -- the hydrodynamic mechanism of the bulk matter
production in $p+p$ collisions, at least, for high multiplicity
events. Then, to reproduce high multiplicity, the initially very
small $p+p$ system has to be superdense at early times. This leads to
very large collective velocity gradients, and so the homogeneity
lengths should be fairly small. However, as we shall demonstrate,
even at the maximally possible velocity gradients at the given
multiplicity, one gets again an overestimate of the interferometry
radii in $p+p$ collisions. The similar result  is obtained in hydrodynamics in Ref. \cite{Bozek}\footnote{The results for $p+p$ \cite{Werner} obtained using EPOS 2.05 + hydro,  need to be clarified since that version of EPOS underestimates the transverse energy per unit of rapidity \cite{Werner}.}.
Therefore, one can conclude that the
problem of theoretical description of the interferometry radii in
$p+p$ collisions may be a general one for different types of event
generators associated with various particle production mechanisms.
Here we try to correct the  results on interferometry from event generators
using for this aim the quantum effects accounting for partial
indistinguishability and mutual coherence of the closely located
emitters due to the uncertainty principle \cite{SinShap}.

In this Letter we employ the hydrokinetic model (HKM)
\cite{HKM, HKM1} in its hybrid form \cite {hHKM} where the UrQMD hadronic
cascade is considered  as the semi-classical event generator at the post
freeze-out (``afterburner'') stage of the hydrodynamic/hydrokinetic evolution. We
analyze two aspects of the analysis of $p+p$
collisions. The main one is:  whether  quantum corrections can help
to describe the experimental data. If yes, it gives hope that it
can be successfully applied for any event generator associated with
another mechanisms of the particle production. The second aspect is
more sophisticated: whether the typical hybrid models developed for
$A+A$ collisions (here hybrid = hydrodynamic/hydrokinetic + hadronic
cascade) with correspondingly modified initial conditions and with the above-mentioned quantum corrections can be real
 candidates to describe the bulk observables in $p+p$
collisions at LHC energies. For this aim we study the space-time
structure of $p+p$ collisions, namely, analyze the multiplicity dependence 
of interferometry radii and volume as well as
the $p_T$-behavior of the HBT radii. It is worth noting that a
satisfactory description of the corresponding experimental data
challenges the theoretical picture of $p+p$ collisions, however supporting
the Landau pioneer suggestion \cite{Landau} to use
relativistic hydrodynamic theory for the hadron collisions with high
multiplicity. Certain arguments in the favor of this suggestion are presented, 
for example, in \cite{SarkSakh1,SarkSakh2}, where multiparticle production in nuclear collisions 
is related to that in hadronic ones within the model based on dissipating energy 
of participants and their types, which includes Landau relativistic hydrodynamics 
and constituent quark picture.

\section{Hydrokinetic model: description and results for $p+p$ collisions}

The hydrokinetic model \cite{HKM, HKM1, hHKM} of $A+A$ collisions consists
of several ingredients describing different stages of the evolution of
matter in such processes.  At the first stage of system's evolution
the matter is supposed to be chemically and thermally equilibrated and
its expansion is described within perfect (2+1)D boost-invariant
relativistic hydrodynamics with the lattice QCD-inspired equation of
state in the quark-gluon phase \cite{qcd} matched with a chemically
equilibrated hadron-resonance gas via crossover-type transition. The
hadron-resonance gas consists of 329 well-established hadron
states\footnote{According to Particle Data Group compilation
  \cite{pdg}.} made of u,d,s-quarks, including $\sigma$-meson
($f_0$(600)).  With such an equilibrated evolution the system reaches
the chemical freeze-out isotherm with the temperature $T_{ch}=$ 165
MeV. At the second stage with $T<T_{ch}$, the hydrodynamically
expanding hadron system gradually looses its (local) thermal and
chemical equilibrium and particles continuously escape from the
system. This stage is described within the hydrokinetic approach
\cite{HKM, HKM1} to the problem of dynamical decoupling. In hHKM model
\cite{hHKM} the hydrokinetic stage is matching with hadron cascade
UrQMD one \cite{urqmd} at the isochronic hypersurface
$\sigma$:~$t=const$ (with $T_{\sigma}(r=0)=T_{ch}$), that guarantees
the correctness of the matching (see \cite{HKM,HKM1, hHKM} for
details). The analysis provided in Ref. \cite{hHKM} shows a fairly
small difference of the one- and two-particle spectra obtained in
hHKM and in the case of the direct matching of hydrodynamics and UrQMD
cascade at the chemical freeze-out hypersurface. Thus, in this Letter
we utilize just the latter simplified ``hybrid'' variant for the
afterburner stage.

Let us try to apply the above hydrokinetic picture to the LHC $p+p$
collisions at $\sqrt{s}=7$ TeV aiming to get the minimal
interferometry radii/volume at the given multiplicity bin. As it is
known \cite{HKM} the maximal average velocity gradient, and so the
minimal homogeneity lengths can be reached for a Gaussian-like
initial energy density profile. For the same aim we use the minimal
transverse scale in ultra-high energy $p+p$ collision, close to the
size of gluon spots \cite{Kopeliovich} in a proton moving with a speed
$v\approx c$. In detail, the initial boost-invariant tube for $p+p$
collisions has a Gaussian energy density distribution in the transverse
plane $\epsilon_i(r)$ with width (rms) $R=$ 0.3 fm \cite{Kopeliovich} and,
following Ref. \cite{hHKM}, we attribute it to an initial proper time
$\tau_0=$ 0.1 fm/c. At this time there is no initial transverse
collective flow. The maximal initial energy density is defined by all
charged particle multiplicity bin.
The maximum initial energy density, $\epsilon_i(r=0)$,  is determined in HKM, 
for selected experimental bins in multiplicity, by fitting of the mean charged particle multiplicity in those bins.

     The correlation function for bosons in the UrQMD event generator is calculated according to
\begin{equation}\label{CFs}
 C(\textbf{q})=\frac{\sum\limits_{i\neq j}\delta_\Delta(\textbf{q}-\textbf{p}_i+\textbf{p}_j)(1+\cos(p_j-p_i)(x_j-x_i))}{\sum\limits_{i\neq j}\delta_\Delta(\textbf{q}-\textbf{p}_i+\textbf{p}_j)}
\end{equation}
where $\delta_{\Delta}(x)=1$ if $|x|<\Delta p/2$ and 0 otherwise,
with $\Delta p$ being the bin size in histograms. The method (\ref{CFs}) accounts for  the smoothness approximation \cite{Pratt4}.  The output UrQMD
3D correlation histograms in the LCMS for different relative momenta
$\textbf{q}=\textbf{p}_1-\textbf{p}_2$ are fitted with  Gaussians
at each $p_T=\frac{\left|{\bf p}_{1T}+{\bf p}_{2T}\right|}{2}$ bin
\begin{equation}
\label{fit}
C(\textbf{q})=1+\lambda \cdot \exp(-R^2_{out}q^2_{out}-R^2_{side}q^2_{side}-R^2_{long}q^2_{long}).
\end{equation}
The interferometry radii $R_{out}(p_T)$, $R_{side}(p_T)$, $R_{long}(p_T)$ and the suppression parameter $\lambda$ are
extracted from this fit.

\begin{figure}
\center
%\vspace{-0.01\textheight}
\includegraphics[height=0.4\textheight]{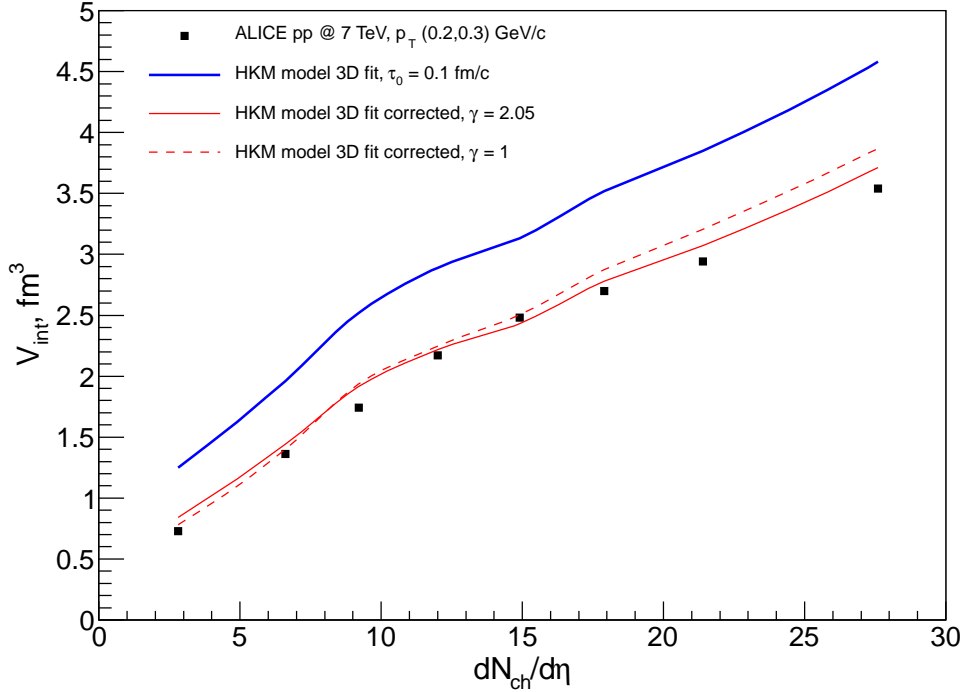}
\caption{The pion
interferometry volume dependency on the charged particles multiplicity
at $p_T=0.2-0.3$~GeV/c. The LHC ALICE data \cite{alice_pp} are compared with
pure HKM results (blue solid line) and with the quantum corrected ones
(red lines).} 
%\vspace{-0.005\textheight}
\label{fig1}
\end{figure}

In Fig.\ref{fig1} we demonstrate  the results from hydrokinetic model for the pion 
interferometric radii, comparing them with the ones measured by the
ALICE Collaboration at the LHC
\cite{alice_pp} in $p+p$ collisions at the energy $\sqrt{s}=7$~TeV.
As one can see there is a significant systematic
overestimate of the predicted interferometry volume $V_{int}=
R_{out}R_{side}R_{long}$ in $p+p$ collision even at the minimal homogeneity lengths
possible for the given multiplicity classes. This  is consistent with
the results of the first paper devoted to the same topic ``Pion
interferometry testing the validity of hydrodynamical models''
\cite{Lorstad}. In what follows we shall try to 
improve the results of the semi-classical HKM event generator by means
of the quantum corrections to them \cite{SinShap}.

\section{The quantum corrections to the hydrokinetic results}
In \cite{SinShap} it is shown that, for small systems formed
in particle collisions (e.g.~$pp$, $e^{+}e^{-}$) where the observed
interferometry radii are about 1--2~fm or  smaller, the uncertainty
principle doesn't allow one to  distinguish completely between
individual emission points. Also the phases of  closely emitted
wave packets are mutually coherent. All that is taken into account in the
formalism of partially coherent phases in the amplitudes of closely
spaced individual emitters. The measure of distinguishability and
partial coherence is then the overlap integral of the two emitted
wave packets. In thermal systems the role of the corresponding coherence
length is played by the thermal de Broglie wavelength that defines also the
size of a single emitter. The Monte-Carlo method (\ref{CFs})
cannot account for such effects since it deals with classical
particles and point-like emitters (points of the particle's last
collision). The classical probabilities are summarized according to
the event generator method (\ref{CFs}), while in the quantum approach
a superposition  of partially coherent amplitudes, associated with
different possible emission points, serves as the input for further
calculations \cite{SinShap}. Such an approach leads to a
reduction of the interferometry radii as compared to 
Eq. (\ref{CFs}). In addition, the ascription of the factor
$1+\cos(x_1-x_2)(p_1-p_2)$ to the weight of the pion pair in
(\ref{CFs})  is not correct for very closely located points $x_1$
and $x_2$ because there is no Bose-Einstein enhancement  if the two
identical bosons are emitted from the same point \cite{Lyub,
SinShap}. The effect is small for large systems with large number of
independent emitters. For small systems, however, it can be significant and one has to exclude 
unphysical contributions (``double counting'' \cite{SinShap}) in the
two-particle emission amplitude. Such corrections lead to a
suppression of the Bose-Einstein correlations that is manifested in a reduction of the observed
correlation function intercept  compared with one in the standard method (\ref{CFs}). 

The results of Ref. \cite{SinShap} are presented in the
non-relativistic approximation related to the rest frame of the
source moving with four-velocity $u^{\mu}$. In the 
hydrodynamic/hydrokinetic approach the role of such a source at a
given pair's half-momentum bin near some value $p$ is played by the fluid element or piece
of the matter with the size equal to the homogeneity length
$\lambda (p)$ \cite{Sin}. These lengths are extracted from the  HKM simulations,
namely, from the interferometry radii defined by the Gaussian fits to
the correlation functions obtained in HKM. All the pairs in procedure
(1) are considered in the longitudinally co-moving system (LCMS) that
in the 
boost-invariant approximation automatically selects the longitudinal
rest frame of the source and longitudinal homogeneity length in this
frame (it is Lorentz-dilated as compared to one in the global system
\cite{Sinyuk}). The femtoscopy analysis is typically related to a fixed
$p_T$ bin and so one needs also to determine the transverse source size in the
transverse rest frame. The corresponding Lorentz transformations
do not change the {\it side}-homogeneity length; as for the  {\it
out}-direction we proceed in the way proposed in Ref.~\cite{Sinyuk}.

Remaining within the  Gaussian approximation, realized for expanding
inhomogeneous systems in the saddle point method \cite{Sin,AkkSin},
let us fix some $\textbf{p} = (\textbf{p}_T,0)$ in the basic LCMS reference system ({\it
basic}-RS) and select the transversely moving reference systems (marked by the
sign {\it tilde}) where the emission density distribution, related to
the space-time center of this local source $\tilde{x}_{0i}(p)$,
$\tilde {t}_{0}(p)$, can be well approximated as the following 
\begin{equation}
\rho(x,t) \propto e^{-{\sum}_{i}\tilde{x}_i^2/2\tilde{\lambda}_{i}^2(p)-\tilde{t}^2/2\tilde{T}^2(p)}
\label{rho}
\end{equation}
Then in {\it
tilde}-RS the correlation function has the form (\ref{fit})
%\begin{equation}
%C(p,\tilde{q})=1+\lambda %e^{-R_{long}^2(p)q_{long}-R_{side}^2(p)q_{side}-\tilde{R}_{out}^2(p) %\tilde{q}_{out}^2}
%\label{CF-base}
%\end{equation}
where $R_{long}^2=\lambda_{long}^2$, $R_{side}^2=\lambda_{side}^2$,
$R_{out}^2\rightarrow
\tilde{R}_{out}^2=\tilde{\lambda}_{out}^2+\frac{\tilde{p}_{out}^2}{\tilde{p}_0^2}\tilde{T}^2$
with $\tilde{T}$ defining the duration of emission in this {\it
  tilde}-RS. The pair's half-momentum $p$ corresponds to the concrete
experimental bin taken in the {\it basic}-RS, the difference of the
particle momenta $q$ components in selected {\it tilde}-RS are
$\tilde{q}_{out}$, $\tilde{q}_{side}=q_{side}$,
$\tilde{q}_{long}=q_{long}$. Therefore, only $q_{out}$ and
correspondingly $R_{out}$, including $\lambda_{out}$ and $T$, are
really transformed in (\ref{fit}) at the Lorentz boosts along the
transverse momentum of the pair. The correlation function $C(p,q)$ is
the Lorentz invariant, therefore $\tilde{R}_{out}^2(p)
\tilde{q}_{out}^2=$ inv.

\begin{figure}
\center
\vspace{-0.02\textheight}
\includegraphics[width=0.95\textwidth]{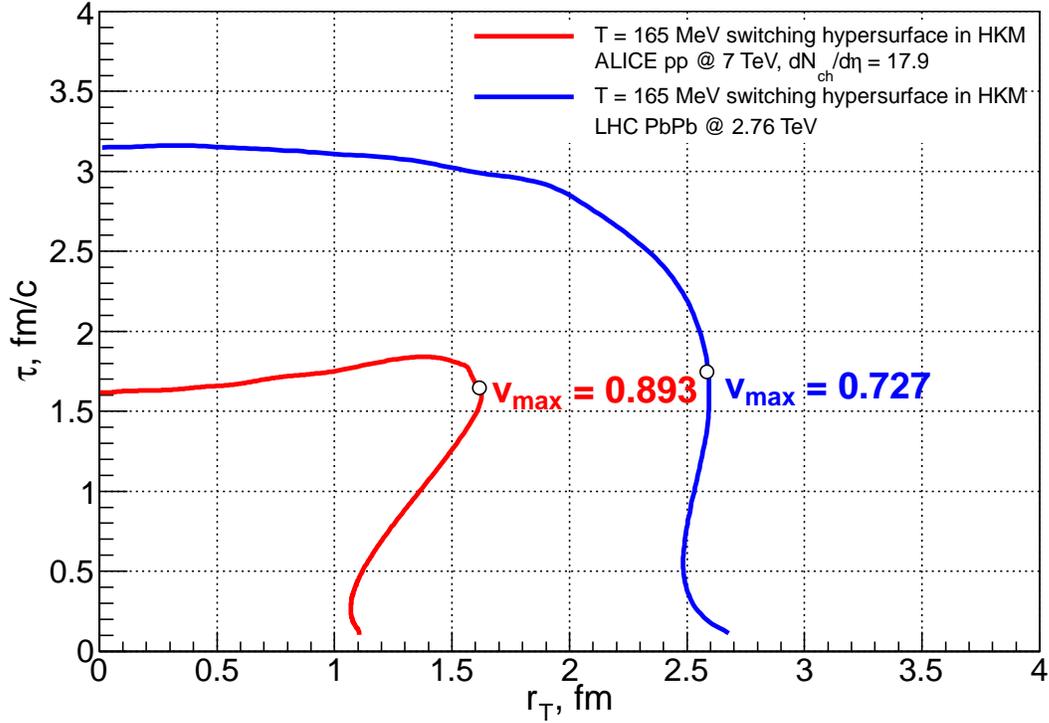}
\caption{The chemical freeze-out hypersurface in HKM in the
transverse plane   for $\sqrt{s}=7$ TeV $p+p$ collisions at
$\frac{dN_{ch}}{d\eta}=17.9$ (red line) in comparison with the
analogous result scaled in both $\tau, r_T$ coordinates by the factor 1/3
for $\sqrt{s}=2.76$ TeV Pb+Pb collisions (blue line). The maximal initial energy densities are close in both cases. The maximal
velocities are marked in the corresponding points on the curves.
} \vspace{-0.01\textheight} \label{fig2}
\end{figure}

To relate the interferometry radius in the rest frame of the source
(marked by the asterisk)  to the one in {\it basic}-RS one should
express  both values through the radius in {\it tilde}-RS using the invariance
property similar as it is done in \cite{Sinyuk}. Then
one can get
\begin{eqnarray}
R^*_{out}(p) &=& R_{out}(p)\frac{\cosh y_T}{\cosh(y_T-\eta_T)}, \,
R^*_{side} = R_{side}, \, R^*_{long}=R_{long} \label{R,T}\\
\lambda^*_{out}&=& \lambda_{out}\frac{\cosh y_T}{\cosh(y_T-\eta_T)}, \,\frac{p^*_{out}}{p^*_0}T^*=
T\frac{\sinh y_T}{\cosh(y_T-\eta_T)} \nonumber
\end{eqnarray}
Here $\eta_T$ is a rapidity of the source in transverse direction,
$y_T=(y_{1T}+y_{2T})/2$ is half-sum of transverse rapidities of the
particles forming the pair. Then one can represent the correlation
function again in the form (\ref{fit}) where all the variables are
related already to the rest frame of the source and the HBT radii in
this rest frame are expressed through the radii in {\it basic}-RS
according to (\ref{R,T}).  Note that $y^*_T=y_T-\eta_T$, and if the
rapidity of the pair is equal to the rapidity of the source,
$y^*_T=0$, then in this particular case the radius in the rest frame
is Lorentz-dilated by the factor $\gamma$. Generally, the reference
system where the pair's momentum is zero does not coincide with the
rest frame of the source that emits the pair. Therefore, the direct
application of these formulas is not an easy task for the rather
complicated emission structure in a hypothetical
hydrodynamic/hydrokinetic model of $p+p$ collisions. In Fig.\ref{fig2}
one can see the structure of the chemical freeze-out hypersurface with the
maximum value of collective velocity labeled for high multiplicity $p+p$
events in comparison with the ones for central Pb+Pb collisions at
LHC\footnote{ Note, that the initial maximal energy densities are
  close in both these processes and the peculiarities of the
  freeze-out hypersurface and velocity profile in $p+p$ case are
  caused by the very large gradients of initial density because of the
  small initial transverse size.}.  Of course, the details of the
transformation will be different for a  string-based  event generator,
therefore we present the analysis for the radii transformation just in
the two limiting cases $R^*_{out}=R_{out}$ and $R^*_{out}=\gamma
R_{out}$ ($\gamma=\cosh y_T$).

We provide the quantum corrections at each $p_T$ bin in the rest frame
of the corresponding source using Eq. (\ref{R,T}) and then come back
again to the {\it basic}-RS. To preserve the previous notations one
can suppose that the source rest frame coincides with {\it
  tilde}-RS. In what follows the {\it tilde} and {\it asterisk} marks
are omitted and all values are related to the source rest frame.  To
account that due to the uncertainty principle the emitters (strictly
speaking emitted wave packets) have finite sizes $\left\langle
  (\Delta x)^2\right\rangle \sim k^{-2}$ ($k$ is the momentum variance of
the particle radiation) when defining the lengths of coherence, one
should at first consider the amplitude of the radiation processes and
only then make statistical averaging over phases of the wave packets
using the overlap integral as the coherence measure \cite{SinShap}.

Following \cite{SinShap} we present the quantum state
$\psi_{x_i}(p,t)$ corresponding to a boson with mass $m$ emitted at
the time $t_i$ from the point ${\bf x}_i$ as a wave packet with
momentum variance $k$ which then propagates freely: 
\begin{equation}
\psi_{x_i}(p,t)=e^{ipx_i -iEt}e^{i\varphi(x_i)}\tilde{f}(\textbf{p})
\label{psi2}
\end{equation}
where $\varphi(x_i)$ is some phase and $\tilde{f}$ defines the primary momentum spectrum $f(\textbf{p})$ that we take in the Gaussian form,
\begin{equation}
f(\textbf{p})=\tilde{f}^2(\textbf{p})=\frac{1}{(2 \pi
k^2)^{3/2}}e^{-\frac{\textbf{p}^2}{2k^2}},
\label{f}
\end{equation}
with the variance $k^2=mT$. The  effective temperature of particle
emission in the local rest frames in HKM, $T$, is close to the
chemical freeze-out temperature $T_{ch}$. 

The amplitude of the single-particle radiation from some 4-volume can
be written at very large times $t_{\infty}$ as a superposition of the
wave functions $\psi_{x_i}(p)$ with  coefficients
$\widehat{\rho}(x_i)=\sqrt{\rho(x_i)}$ that leads in the case of
completely random phases to the emitter distribution (\ref{rho}) in
the local rest frames of the sources: 
\begin{equation}
A(p,t)=c\int d^4x_i \psi_{x_i}(p,t)
\widehat{\rho}(x_i),
\label{A}
\end{equation}
where $c$ is the normalization constant.

In the  paper \cite{SinShap}  the two-particle state is considered  as a product of the single-particle amplitudes,  thus suggesting the  maximal possible distinguishability and independence of different emitters compatible with the uncertainty principle for momentum \& position and energy-momentum \& time  measurements. The latter is accounted for by the averaging of such a two-particle amplitude over  partially coherent phases between different emitters with overlap integral measure \cite{Tolst, SinShap} 
\footnote{Such a  'minimal'  consideration of the uncertainty principle only does not exclude,  of course,  an existence of the concrete mechanisms of correlation and coherence between emitters; note, however, that  such more complicated  picture might lead to the results different from these that are set forth here.}. 

With that said the single- and two-particle spectra, averaged over
the ensemble of emission events with partially correlated
phases $\varphi(x)$ are
\begin{eqnarray}
\overline{W(p)} & =&  c^2 \int d^4x d^4x^{\prime} e^{ip(x-x^{\prime})}\widehat{\rho}(x)
\widehat{\rho}(x^{\prime}) f(\textbf{p}) \langle e^{i(\varphi(x)-\varphi(x^{\prime}))} \rangle  \nonumber\\
\overline{W(p_1,p_2)} &=&  c^4\int d^4x_1 d^4 x_2 d^4 x_1^{\prime} d^4 x_2^{\prime}e^{i(p_1
x_1+ p_2 x_2 -p_1 x_1^{\prime} - p_2x_2^{\prime})} \cdot \nonumber\\
&& \cdot f(\textbf{p}_1) f(\textbf{p}_2)
\widehat{\rho}(x_1)\widehat{\rho}(x_2)\widehat{\rho}(x_1^{\prime})\widehat{\rho}(x_2^{\prime}) \langle
e^{i(\varphi(x_1)+\varphi(x_2)-\varphi(x_1^{\prime})-\varphi(x_2^{\prime}))}\rangle.
\end{eqnarray}
The phase averages are associated with corresponding overlap integrals \cite{SinShap}
\begin{eqnarray}
 \langle e^{i(\varphi(x)-\varphi(x^{\prime}))} \rangle &=&
 G_{xx^{\prime}} = I_{xx^{\prime}}= \left|\int d^3{\bf
     r}\psi_{x}(t,{\bf r})\psi_{x^{\prime}}^*(t,{\bf r})\right|, \\ 
 \langle
 e^{i(\varphi(x_1)+\varphi(x_2)-\varphi(x_1^{\prime})-\varphi(x_2^{\prime}))}\rangle
 &=&
 G_{x_1x_1^{\prime}}G_{x_2x_2^{\prime}}+G_{x_1x_2^{\prime}}G_{x_2x_1^{\prime}}-G_{x_1x_2^{\prime}}G_{x_2x_1^{\prime}}G_{x_1x_2} 
\end{eqnarray}
where $\psi_{x_i}(t,{\bf r})=\frac{1}{(2\pi)^{3/2}}\int
f(\textbf{p})e^{-i {\bf p (r-x}_i)}e^{-i\frac{{\bf
      p}^2}{2m}(t_i-t)}d^3p$ 
 \, are the wave functions of single bosonic states in coordinate representation. 

Then the correlation function $C({\bf p},{\bf q})$ can be expressed
through the homogeneity lengths in the local rest frame
$R_{L}\equiv\lambda_{long}^*(p)$, $R_{S}\equiv\lambda_{side}^*(p)$,
$R_{O}\equiv\lambda_{out}^*(p)$ that are expressed through the HBT
radii obtained from the Gaussian fit (\ref{fit}) of the HKM
correlation functions and transformation law (\ref{R,T}) as described
above. 
\begin{eqnarray}
&&C({\bf p},{\bf q})=\frac{\overline{W(p_1,p_2)}}{\overline{W(p_1)}\overline{W(p_2)}} =\nonumber \\
&=&1+e^{- q_O^2R_O^2 \frac{4k_0^2R_O^2}{1+4k_0^2 R_O^2}
- q_S^2 R_S^2 \frac{4k_0^2R_S^2}{1+4k_0^2 R_S^2}
-q_L^2R_L^2 \frac{4k_0^2R_L^2}{1+4k_0^2 R_L^2}-\frac{({\bf q}\cdot {\bf p})^2T^2}{m^2} \frac{4k^2T^2}{1+4k^2T^2}} - C_d({\bf p},{\bf q}) ,
\label{corf}
\end{eqnarray}
where $k_0^2=k^2/(1+ \alpha k^4 T^2/m^2)$, parameter $\alpha(k^2R^2)$
is defined from the model numerically (it is the order of unity for
$R\sim 1$ fm and tends to zero for the large sources -- see
\cite{SinShap} for details), and the subtracted term 
\begin{eqnarray}
C_d({\bf p},{\bf q}) = e^{-\frac{2 q_O^2 k_0^2 R_O^4  \left(1+8 k_0^2 R_O^2\right)}{\left(1+4 k_0^2 R_O^2\right) \left(1+8 k_0^2R_O^2+8k_0^4R_O^4 \right)}
-\frac{2 q_S^2 k_0^2 R_S^4  \left(1+8 k_0^2 R_S^2\right)}{\left(1+4 k_0^2 R_S^2\right) \left(1+8 k_0^2R_S^2+8k_0^4R_S^4 \right)}
-\frac{2 q_L^2 k_0^2 R_L^4  \left(1+8 k_0^2 R_L^2\right)}{\left(1+4 k_0^2 R_L^2\right) \left(1+8 k_0^2R_L^2+8k_0^4R_L^4 \right)}} \cdot \nonumber \\
\cdot e^{-\frac{2 k^2 T^4 ({\bf p} \cdot {\bf q})^2 \left(1+8 k^2 T^2\right)}{m^2 \left(1+4 p^2 T^2\right) \left(1+8 k^2
   T^2+8 k^4 T^4\right)}} F(k_0^2R_i^2,k^2T^2), \nonumber
\end{eqnarray}
\begin{eqnarray}
F(k_0^2R_i^2,k^2T^2)  = \left(\frac{k_0}{k} \right)^{3/2} \left(
\frac{1+4 k^2 T^2}{1+8 k^2 T^2+8 k^4 T^4}
\frac{1+4 k_0^2 R_O^2}{1+8 k_0^2 R_O^2+8 k_0^4 R_O^4} \cdot \right. \nonumber \\
\left. \cdot \frac{1+4 k_0^2 R_S^2}{ 1+8 k_0^2 R_S^2+8 k_0^4 R_S^4}
\frac{1+4 k_0^2 R_L^2}{1+8 k_0^2 R_L^2+8 k_0^4 R_L^4}\right)^{1/2}
\end{eqnarray}
corresponds to the elimination of the double counting.

Now we can see that the apparent interferometry radii extracted from the Gaussian fits to the correlation function (\ref{corf})
are reduced as compared to those obtained in the standard approach.

Particularly, if we  neglect the
double counting effects, truncate the subtracted term $C_d({\bf p},{\bf q})$ in (\ref{corf}), and fit the correlation
function with the Gaussian (\ref{fit}), we obtain the femtoscopic radii $R_{out}$, $R_{side}$, $R_{long}$
related to the standard ones $R_{out,st}$, $R_{side,st}$, $R_{long,st}$ as follows
\begin{eqnarray}
\frac{R^2_{out}}{R_{out,st}^2} & = &
\left(R_O^2\frac{4k_0^2R_O^2}{1+4k_0^2
    R_O^2}+T^2v^2_{out}\frac{4k^2T^2}{1+4k^2T^2}\right)/\left(R_O^2+T^2v^2_{out}\right)
\nonumber \\ 
\frac{R^2_{side}}{R_{side,st}^2} & =& \frac{4k_0^2R_S^2}{1+4k_0^2 R_S^2}  \\
\frac{R^2_{long}}{R_{long,st}^2} & = &\frac{4k_0^2R_L^2}{1+4k_0^2 R_L^2} \nonumber
\end{eqnarray}
where $v_{out}= p^{*}_{out}/p_0^* \ll 1$ according to the
non-relativistic approximation. For large source sizes, e.g. when the
homogeneity lengths correspond to $A+A$ collisions, $k_0^2R^2 \gg 1$,
$k^2T^2 \gg 1$, all these ratios tend to unity. 

The mean emission duration is supposed to be proportional to the
average system size, $T = a (R_O+R_S+R_L)/3$ that leads to a
quadratic equation expressing $R_O$ (and $T$) through $R_{i,st}$. The
latter are connected with ones taken in the {\it basic}-RS according
to transformation laws (\ref{R,T}).  The value $a$ is a free model
parameter.  Then we put these extracted values into the expression
(\ref{corf}) for the correlation function and perform its fitting with
the Gaussian (\ref{fit}). This gives us finally the interferometry
radii $R_{out}$, $R_{side}$ and $R_{long}$ in view of the uncertainty
principle. The radii are presented then in the {\it basic}-RS using
the transformations inverse to~(\ref{R,T}).

The correlation function is the ratio of the two- and one-particle
spectra. It is found~\cite{SinShap} that quantum corrections to this
ratio are not so sensitive to different forms of the wave packets as
the spectra itself. In particular, the effective temperature of the
{\it corrected} transverse spectra depends on whether the parameter of
mean particle momentum is included or not into the wave packet
formalism. If yes, the corrected effective temperature for small
sources $R\sim 1$ fm is equal or even higher than that of  individual
emitters, $T=k^2/m$, while for the wave packets in the form
(\ref{psi2}) it is lower \cite{SinShap}. Besides of this, in the
non-relativistic approximation one can describe only very soft part of
the spectra.  That is why we focus in the Letter on the corrections
to the Bose-Einstein correlation functions where in the rest frame of
the source the total and relative momenta of the boson pairs are
fairly small.

\section{The results for $p+p$ and $p+\textrm{Pb}$ collisions, and discussion}

The initial conditions for HKM are described in Section 2. The HKM
event generator provides us with the interferometry radii in {\it
  basic}-RS. To find the corresponding homogeneity lengths in the rest
frame of the source according to (\ref{R,T}) we use, as discussed in
Section 3, the two limiting cases: the transverse boost to the rest
frame of the pair from the basic LCMS system, or no transformation at
all. For the former case it is defined by the $p_T$ bin and for
$p_T=0.2 - 0.3$ GeV $\cosh y_T =\gamma=2.05$. The parameter $a$
connecting $T$ with $R_i$ increases linearly with multiplicity from
0.8 to 1.0, primary momentum spectrum dispersion $k=0.16$~GeV/c
(T=0.18 GeV), pair mean transverse momentum in the source rest frame
$p^{*}_{T}=0.15$~GeV/c. The $\alpha$ parameter is set to linearly
decrease with multiplicity from 0.8 to 0.6.  As for the $\gamma=1$ case,
the parameter $a$ decreases linearly with multiplicity from 1.1 to 0.9,
$k=0.18$~GeV/c and $p^{*}_{T}=0.13$~GeV/c. The $\alpha$ parameter
decreases linearly with multiplicity from 1.35 to 0.9 which is close to the
theoretical results \cite{SinShap}.
\begin{figure}
\center
%\vspace{-0.02\textheight}
\includegraphics[width=0.95\textwidth]{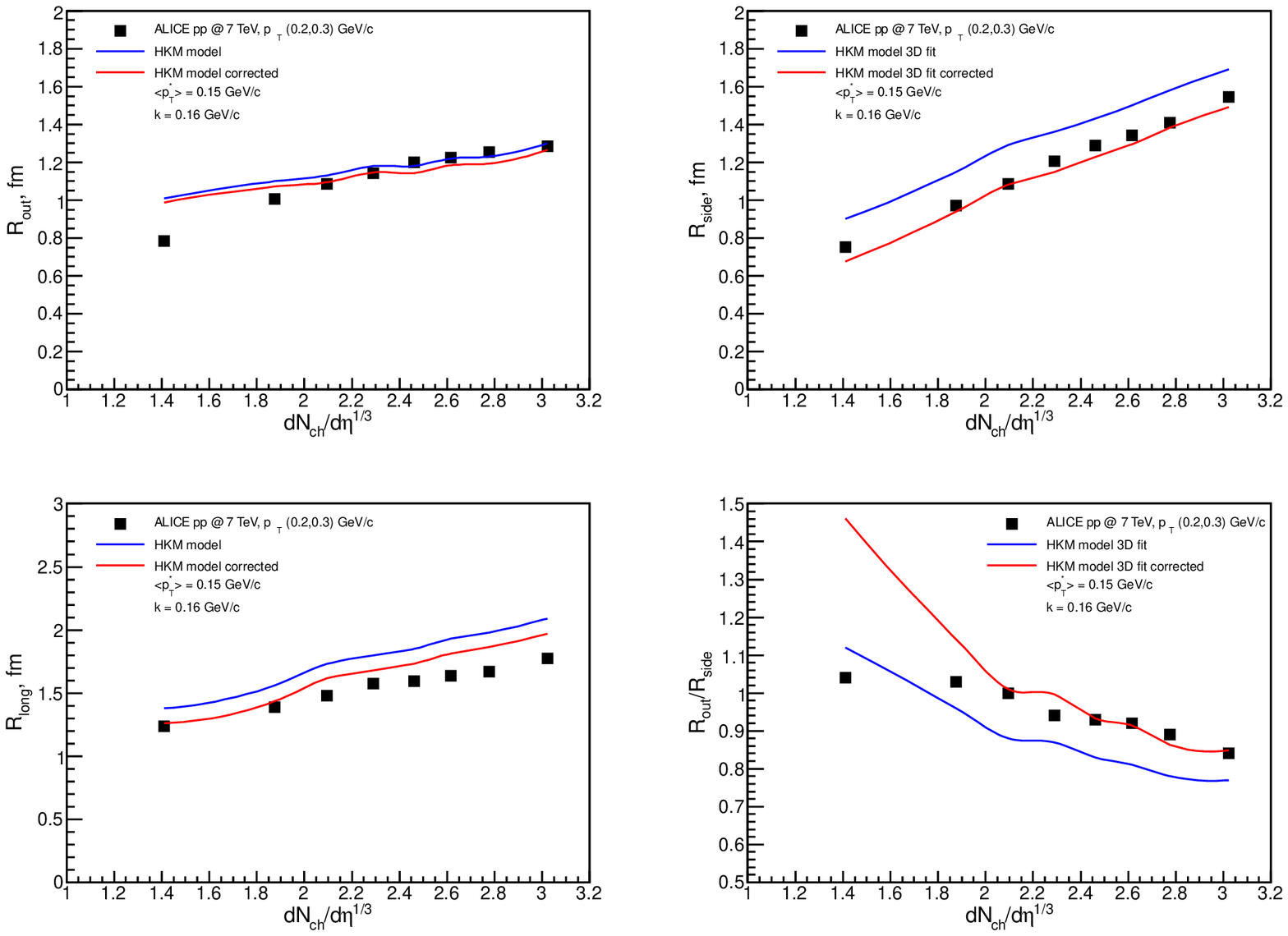}
\caption{The pion interferometry radii dependency on charged particles
  multiplicity} at $p_T=0.2-0.3$~GeV/c. The designations are the same as
in Fig. \ref{fig1}. 
\vspace{-0.01\textheight}
\label{fig3}
\end{figure}
In Fig. \ref{fig1} along with the experimental and pure HKM results we
present the multiplicity dependence of the quantum corrected
interferometry volume at $p_T=0.25$ GeV/c.  The solid line represents
the corrected values calculated under the assumption that the $R_{out}$
interferometry radii, observed in {\it basic}-RS, are
Lorentz-contracted by a factor $\gamma = 2.05$ for the chosen $p_T=0.25$
GeV/c value as compared to ones in the source rest system. The dashed line
demonstrates the no-contraction case when $\gamma = 1$.  As one can
see the accounting for the uncertainty principle allows one to
describe the overall multiplicity dependence of the interferometric radii. 
Figure \ref{fig3} represents the dependence on multiplicity of
individual radius parameters.

The suppression of the Bose-Einstein correlations for small sources
with closely located emitters takes place even without specific
coherence mechanism and resonance contributions. To see this effect the
double counting in the correlation function should be eliminated as
Eq.~(\ref{corf}) demonstrates. Then the additional suppression
parameter $\lambda_{\text coh} < 1$ in the Gaussian fit appears and
the relevant parameter in (\ref{fit}) becomes $\lambda =\lambda_{\text
  coh}\lambda_\text{HKM}$.  The result of our calculations gives
$\lambda_{coh}=$ 0.9~--~0.95 for not very small multiplicities.

\begin{figure}
\center
%\vspace{-0.02\textheight}
\includegraphics[height=0.4\textheight]{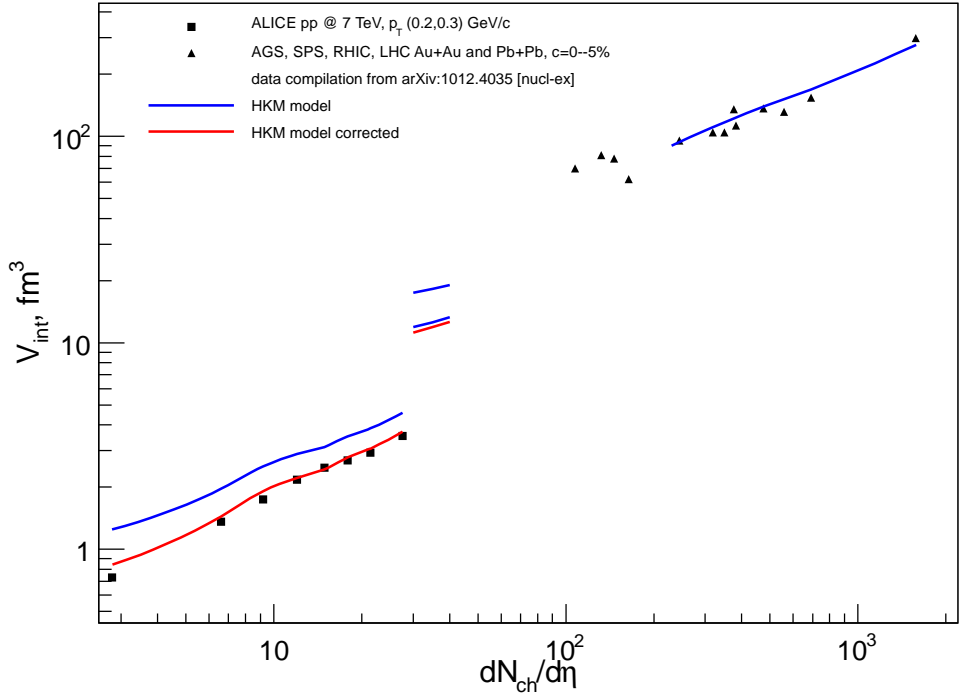}
\caption{The interferometry volume dependency on charged particles
  multiplicity. The curve fragments in the middle correspond to the
  HKM prediction for $p+$Pb collision at the LHC energy
  $\sqrt{s}=5.02$~GeV. The upper one is related to the initial
  transverse system size $R=1.5$~fm and for the two lower ones
  $R=0.9$~fm.  The curves at the left and at the right represent the
  HKM results for $p+p$ and $A+A$ central collisions respectively,
  compared to the experimental data at AGS, SPS, RHIC and LHC, taken
  from papers \cite{alice_pp}, \cite{AaExp1} -- \cite{AaExp8}. The
  $pp$ volumes are calculated as a product $R_{out}R_{side}R_{long}$
  of respective experimental radii. The blue lines correspond to pure
  HKM results, whereas the quantum corrections to them are presented
  by the red lines.}
\vspace{-0.01\textheight}
\label{fig3a}
\end{figure}

In addition to the correlation analysis of $p+p$ collisions, let us
make the simplest estimates and try to predict the HBT radii
for $p+$Pb collisions at the LHC energy $\sqrt{s}=5.02$~GeV. We ignore
the possible asymmetry of the hydrodynamic tube in the longitudinal
direction and present our prediction within hHKM for centrality
$c=0-20$~\% with $dN_{ch}/d\eta$ = 35. The results are calculated for
the two initial radii with rms equal to 0.9 fm and 1.5 fm and also for
the two initial times: $\tau =$ 0.1 fm/c and 0.25 fm/c. It turns out
that the latter factor is not essential if we keep fixed final
multiplicity: only the longitudinal radii are 3--4\% higher at $\tau
=$ 0.25 fm/c than at 0.1 fm/c. The transverse radii practically
coincide. Therefore, we finally demonstrate only the case $\tau =$ 0.1
fm/c.  The initial transverse sizes of the system, created in the
$p+$Pb collision, are taken from Ref. \cite{Boz}: ``In the
conventional wounded nucleon model it is assumed that the sources are
located in the transverse plane in the centers of the participating
nucleons. This amounts to rather large initial transverse sizes in the
$p-$Pb system, $R=1.5$~fm. Locating the source in the center-of-mass
of the $NN$ system is also admissible, which leads to a more compact
initial distribution, $R=0.9$~fm''. The results for the interferometry
volume are presented in Fig.~\ref{fig3a}. The model parameter set
is extrapolated from the described above and consistent with that for the $p+p$
system, with $\gamma = 1$ to the case of larger sizes
typical for the $p+$Pb collisions. At that the $k$ and $p^{*}_{T}$
values are left the same as for the $p+p$ case, whereas $\alpha$ and
$a$ are chosen to be smaller. For the $R=0.9$~fm initial transverse
size $\alpha=0.5$, $a=0.7$ and for $R=1.5$~fm we put $\alpha=0.45$,
$a=0.6$.

Considering the multiplicity dependence of femtoscopy scales in $p+p$ and $p+$Pb collisions we cannot bypass the  scaling
hypothesis issue \cite{Lisa}, that suggests a universal linear dependence of the HBT volume on the particle multiplicity. It means that the observed interferometry volume
depends roughly only on the multiplicity of particles produced in collision, but not on the geometrical characteristics of the collision process. At the same time, as it was found in the theoretical analysis in Ref. \cite{scaling}, the interferometry volume should depend not only on
the multiplicity, but also on the initial size of colliding systems. In more detail, the intensity of the transverse flow depends on the initial geometrical size $R^g_0$ of the system: roughly, if the pressure is  $p = c_0^2\epsilon$, then the transverse acceleration $a= \nabla_{x_T} p/\epsilon \propto p({\bf x}_T= 0)/(R_0^g\epsilon)=c_0^2/R^g_0$. The interferometry radii $R_T$, that are associated with the homogeneity lengths, depend on the velocity gradient and geometrical size,   and  for non-relativistic transverse expansion can be approximately expressed through $R^g_0$, the averaged transverse velocity $\left\langle |v_T|\right\rangle$  and inverse of the temperature $\beta$ at some final moment $\tau$ \cite{AkkSin, Csorgo, PBM}: 
\begin{equation}
R_T= \frac{R^g(\tau)}{\sqrt{1+\frac{2}{\pi} \langle |v_T|\rangle^2\beta m_{T}}}\approx 
R_0^g \left(1+\frac{\tau^2 c_0^2}{2\left(R^g_0\right)^2}-\beta m_T\frac{\tau^2c^2_0}{\pi^2 \left(R^g_0\right)^2}\right)
\label{compensation}
\end{equation} 
The result (\ref{compensation}) for the HBT radii  depends obviously on $R_0^g$ and, despite its roughness, demonstrates  the possible mechanism of compensation of the growing (in time) geometrical radii of an expanding fireball in the femtoscopy measurements. For some dynamical models of expanding fireballs \cite{Csorgo2} the interferometry radii, measured at the final time of system's decoupling, are fully coincided with the initial geometrical ones, no matter how large the multiplicity is. The reason for such a behavior is explained in Ref. \cite{HKM}: if there is no dissipation in the expanding system, namely, the evolution corresponds to a solution of the Boltzmann equation with $F^{gain}(t,{\bf x}) = F^{loss}(t,{\bf x})$, then the spectra and correlation functions are coincided with the initial ones. The detail study of hydrodynamically expanding systems is provided in Ref. \cite{scaling}. It is found that at the boost-invariant isentropic and chemically frozen evolution the interferometry volume, if it were possible to measure the interferometry radii at some evolution time $\tau$, is approximately constant:       
\begin{equation}
V_{int}(\tau) \simeq C(\sqrt{s})\frac{dN/dy(\tau)}{\langle f\rangle_{\tau}
T_{eff}^{3}(\tau)}
\end{equation}
where $\left\langle f\right\rangle$ is the averaged phase-space density \cite{Bertsch} which is found to be approximately conserved during the hydrodynamic evolution under above conditions  as well as $\frac{dN}{dy}$ \cite{scaling}. As for the effective temperature of the hadron spectra, $T_{eff}(\tau)=T(\tau)+m\frac{\left\langle v_T(\tau)\right\rangle^2 }{2}$, one can see that when the system's temperature $T$ drops, the mean $v_T^2$ increases, therefore $T_{eff}$ does not change much during the evolution (it slightly decreases with time for pions and increases for protons). Hence $V_{int}$,  if it has been measured at some evolution time $\tau$, will also approximately conserve. Of course, the real evolution is neither isentropic, nor chemically frozen, includes also  QGP stage, but significant dependence of the femtoscopy scales on the initial system size is preserved anyway.

Fig. \ref{fig3a}
shows the dependency $V_{int}(\langle dN_{ch}/d\eta \rangle)$ for
the case of $p+p$ collisions at the~LHC, $\sqrt{s}=7$~TeV, and
for the most central (only!) collisions of nuclei having similar
sizes, Pb+Pb and Au+Au, at the SPS, RHIC and LHC. We have also added
on the plot our prediction for the interferometry volume of
$p$Pb system, that has an initial size larger than that for the $pp$ system.  As one
can see, the different groups of points corresponding to $p+p$, $p+$Pb
and $A+A$ events cannot be fitted by the same straight line. This
apparently confirms the result obtained in \cite{scaling} that the
interferometry volume is a function of both variables: the
multiplicity and the initial size of colliding system. The latter
depends on the atomic number $A$ of colliding objects and the
collision centrality $c$.

\begin{figure}
\center
\vspace{-0.01\textheight}
\includegraphics[width=0.88\textwidth]{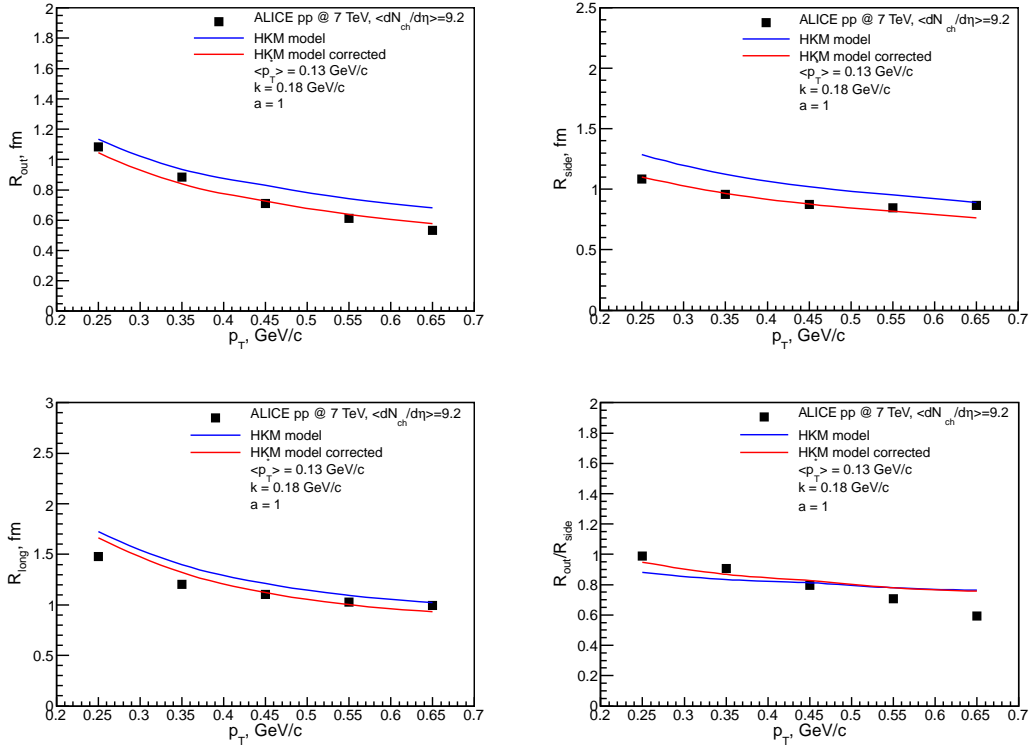}
\caption{Dependence of interferometry radii  on $p_T$, $\langle dN_{ch}/d\eta \rangle = 9.2$}
\vspace{-0.005\textheight}
\label{fig4}
\end{figure}

\begin{figure}
\center
%\vspace{-0.02\textheight}
\includegraphics[width=0.88\textwidth]{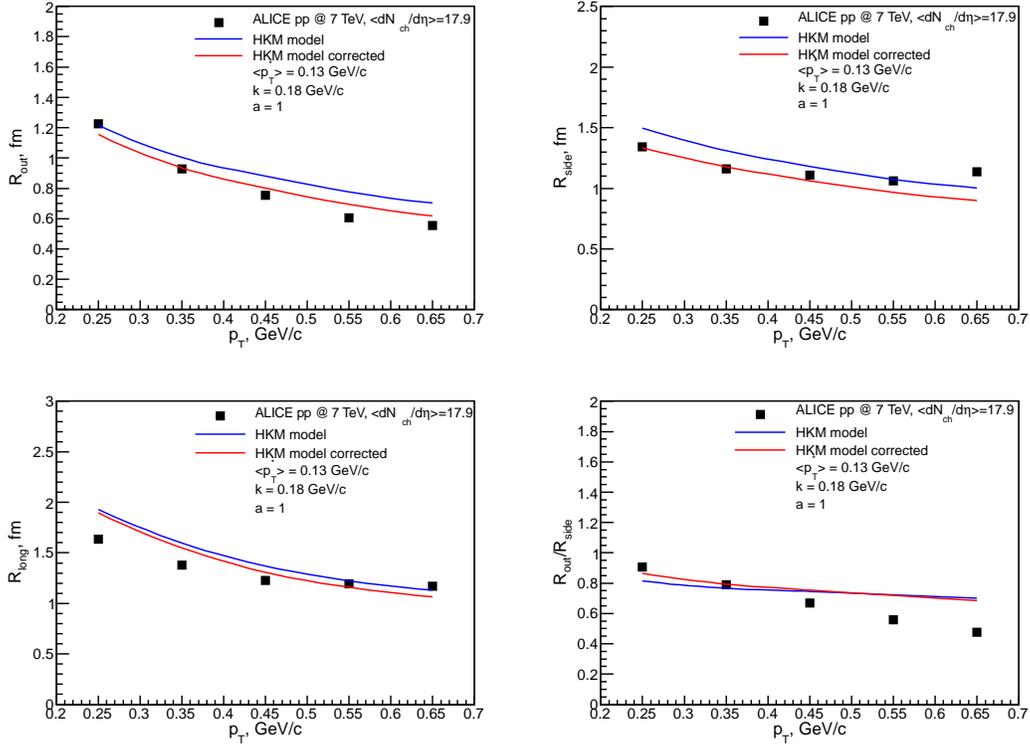}
\caption{Dependence of interferometry radii  on $p_T$, $\langle dN_{ch}/d\eta \rangle = 17.9$}
\vspace{-0.02\textheight}
\label{fig5}
\end{figure}

In Figs. \ref{fig4} and \ref{fig5} for the two multiplicity
classes $\langle dN_{ch}/d\eta \rangle = 9.2$ and $\langle
dN_{ch}/d\eta \rangle = 17.9$ we present the three curves for
interferometry radii as a function of $p_T$: the experimental one, the
one taken just from the HKM simulations and the other one obtained
after application of the quantum corrections.  The basic parameters
used correspond to the case $\gamma=1$ (see above). The
$\alpha$~parameter values linearly increase with $p_T$ from 1.15 to 1.35
for the $\langle dN_{ch}/d\eta \rangle = 9.2$ case and from 1.02 to 1.10
for $\langle dN_{ch}/d\eta \rangle = 17.9$ in such a way that for
$p_T$ bin (0.2,0.3) it has the same values as in previous $R_i(\langle
dN_{ch}/d\eta \rangle)$ calculations.  As one can see, similarly to
the multiplicity behavior, the quantum corrected $p_T$-dependency of
the radii also gets closer to the experimental values, but for large
$p_T$ the corrections are insufficient to fully describe the
observable femtoscopy scales behavior.  This fact may indicate that
sources of particles with large $p_T$ cannot be described in
hydrodynamical approximation. Note that just for such large $p_T$ the
non-trivial base-line corrections, already provided in presented
experimental data, are very essential.

\begin{figure}
\center
%\vspace{-0.02\textheight}
\includegraphics[width=0.88\textwidth]{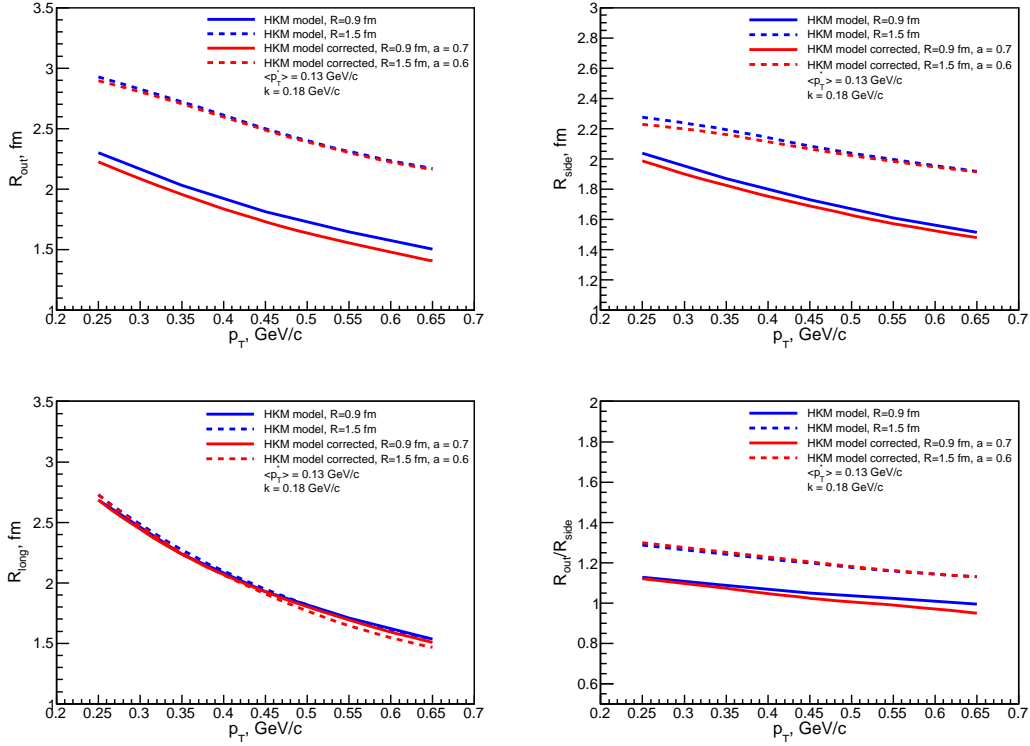}
\caption{The HKM prediction for the dependence of $p+$Pb interferometry radii 
  on $p_T$ at the LHC energy $\sqrt{s}=5.02$~GeV, $\langle
  dN_{ch}/d\eta \rangle = 35$.} 
%\vspace{-0.02\textheight}
\label{pPb}
\end{figure}

Finally we demonstrate the prediction of the $p_T$-dependence of the
radii for $p+$Pb collisions. It is presented in Fig.\ref{pPb}
together with 
the corrections due to the uncertainty principle. One can see that
corrections are smaller for the systems with larger homogeneity lengths
and not very essential for $p+$Pb and probably for $A+A$ peripheral
collisions. As for the homogeneity lengths formation, in a very recent paper \cite{Bzdak} 
it is found that, in the absence of hydrodynamic flow, the HBT radii should be similar in $pp$ and $p$Pb collisions.

\section{Conclusions}
One can conclude that quantum corrections to pion interferometry
radii in $p+p$ collisions at the LHC can significantly improve the
(semi-classical) event generator results that typically give an
overestimate of the experimental interferometry radii and volumes. The
corrections account for the basic (partial) indistinguishability and
mutual coherence of closely located emitters because of the
uncertainty principle \cite{SinShap}. The additional suppression of
the Bose-Einstein correlation function also appears. The effects
become important for small sources, 1--2 fm or smaller. Such systems
cannot be completely random and so require a modification of the
standard theoretical approach for correlation analysis. The
predicted interferometric radii for $p+$Pb collisions need some small
corrections only for its minimal values corresponding to the {\it
  initial} transverse size of $p$Pb system 0.9 fm.

More sophisticated result of this study is a good applicability of the
hydrodynamics/hydrokinetics with the quantum corrections for
description of HBT radii not only in $A+A$ collisions but
also, at least for large multiplicities, in $p+p$ events. These radii
are well reproduced for not too large $p_T$. Whether it means the validity
of the hydrodynamic mechanism for the bulk matter production in the
LHC $p+p$ collisions is still an open question. It is also related to
the problem of early thermalization in the processes of heavy ion
collisions; the nature of this  phenomenon is still a fundamental
theoretical issue.

\section{Acknowledgment}
Yu.M.S. is grateful to B. Kopeliovich, R. Lednick\'y, L.V. Malinina,
E.E. Zabrodin for fruitful discussions and to the ExtreMe Matter Institute
EMMI for visiting professor position.  Iu.A.K. acknowledges the
financial support by the ExtreMe Matter Institute EMMI and Hessian
LOEWE initiative. The research was carried out within the scope of the
EUREA: European Ultra Relativistic Energies Agreement (European
Research Group: ``Heavy ions at ultrarelativistic energies''), and is
supported by the State Fund for Fundamental Researches of Ukraine
(Agreement F33/24-2013).

\end{document}